\documentclass[10pt,preprint]{revtex4-1}
\usepackage{graphicx}
\usepackage{hyperref}
\usepackage{amsmath,amssymb,dsfont,mathrsfs,mathtools}
\usepackage{enumerate}
\usepackage{xcolor}

\newcommand{\Pe}{\mathbb{P}} 
\newcommand{\A}{\mathbb{A}} 
\newcommand{\V}{\mathbb{V}} 
\newcommand{\C}{\mathbb{C}} 
\newcommand{\p}{\partial}
\newcommand{\Tr}{\text{Tr}}

\newcommand{\RR}{\mathds{R}}

\newcommand{\EE}{\mathbb{E}}

\newcommand{\eps}{\varepsilon}

\renewcommand{\Re}{R\mspace{-2mu}e}

\newcommand{\nn}{\nonumber}
\begin{document}
	
\title{Extreme events and instantons in Lagrangian passive scalar
  turbulence models}

\author{Mnerh \surname{Alqahtani}}
\affiliation{Mathematics Institute, University of Warwick, Coventry
  CV4 7AL, United Kingdom}

\author{Leonardo \surname{Grigorio}}
\affiliation{Centro Federal de Educação Tecnológica Celso Suckow da Fonseca -- RJ}

\author{Tobias \surname{Grafke}}
\affiliation{Mathematics Institute, University of Warwick, Coventry
  CV4 7AL, United Kingdom}
\date{\today}

\begin{abstract}
  The advection and mixing of a scalar quantity by fluid flow is an
  important problem in engineering and natural sciences. If the fluid
  is turbulent, the statistics of the passive scalar exhibit complex
  behavior. This paper is concerned with two Lagrangian scalar
  turbulence models based on the recent fluid deformation model that
  can be shown to reproduce the statistics of passive scalar
  turbulence for a range of Reynolds numbers. For these models, we
  demonstrate how events of extreme passive scalar gradients can be
  recovered by computing the instanton, i.e., the saddle-point
  configuration of the associated stochastic field theory. It allows
  us to both reproduce the heavy-tailed statistics associated with
  passive scalar turbulence, and recover the most likely mechanism
  leading to such extreme events. We further demonstrate that events
  of large negative strain in these models undergo spontaneous
  symmetry breaking.
\end{abstract}

\maketitle


\section{Introduction}

The 3D incompressible \emph{Navier-Stokes equations} (NSE),
\begin{equation}
  \partial_t u + u\cdot\nabla u + \nabla p - \nu \Delta u = 0, \qquad \nabla \cdot u = 0, \label{eq:NSE}          
\end{equation} 
describe the evolution of a fluid in time. Here, $u\left(x, t\right)
\in \mathbb{R}^3$ is the velocity field, $\nu$ denotes the kinematic
viscosity, and $p(x,t)\in\RR$ is the scalar pressure field that
enforces the incompressibility constraint. A passive scalar, such as a
substance concentration (e.g., pollutant or temperature field without
a back reaction), is advected by a turbulent flow exhibiting complex
spatial and temporal scales of motions.  The \emph{passive scalar
  equation} (PSE) gives its time evolution,
\begin{equation}  
  \partial_t \theta + u \cdot \nabla \theta - \kappa \, \Delta \theta = 0,    \label{eq:PS}
\end{equation}
where $\kappa$ denotes the diffusivity coefficient of $\theta\left(x,
t\right) \in \mathbb{R}$.  Passive scalar turbulence is often taken as
a testbed for understanding fluid
turbulence~\cite{shraiman-siggia:2000,
  falkovich-gawedzki-vergassola:2001}, but is also relevant in its own
right to analyze, for example, advection processes in the atmosphere
or ocean.

Understanding the statistical and geometrical properties of turbulent
flow at small scales has been a long-standing challenge. At these
scales of motion, the prolific activity of strain and vorticity
triggers intense fluctuations, resulting in \emph{intermittency}, as
observed in the probability distribution functions (PDFs) of velocity
gradients~\cite{frisch:1995}. The velocity gradient not only dominates
the smallest scales of motion, but it also embodies local rotation and
deformation rate, making it an observable object of theoretical
\cite{vieillefosse:1982, vieillefosse:1984, cantwell:1992,
  meneveau:2011} and numerical/experimental studies
\cite{wallace:2009}.
	
Aiming at obtaining the statistics of the small scales provided by the
velocity gradients $A_{ij} = \p u_i/\p x_j$, a variety of low
dimensional models has been proposed in the literature describing the
evolution of $A_{ij}$ following a tracer particle (Lagrangian
description). As the effect of pressure and viscosity renders the
dynamical equation for $A_{ij}$ unclosed, one is forced to resort to
some closure approximation to obtain a self-contained
model~\cite{girimaji-pope:1990}. The \emph{restricted Euler} (RE)
equation \cite{vieillefosse:1982}, the \emph{tetrad model}
\cite{chertkov-pumir-shraiman:1999}, and the \emph{recent fluid
  deformation} (RFD) \cite{chevillard-meneveau:2006} form a history of
such models, where in particular the last has successfully regularized
the finite time singularity of the nonlinear self-stretching term ($-
\A^2$) observed in the RE model, using ideas from linear damping
\cite{girimaji-pope:1990,jeong-girimaji:2003} and geometrical
considerations of \cite{chertkov-pumir-shraiman:1999}. At the same
time, it preserves the statistical features of the velocity gradient
such as the left-skewness of its distribution, and the properties of
the joint PDFs in the $Q$-$R$ plane, where $Q$ and $R$ are the first
and second invariants of $\A$, respectively.
	
With similar arguments, the Lagrangian evolution of a passive scalar
can be added. The resulting passive scalar RFD model (PS-RFD) proposed
in~\cite{gonzalez:2009} retains the statistical properties of the
scalar gradient ${\psi \coloneqq \nabla \theta}$, such as the
stretched exponential PDFs of $\psi$ deviating from Gaussian at small
scales, in excellent agreement with full direct numerical simulations
of passive scalar turbulence~\cite{hater-homann-grauer:2011}. Extreme
values of the scalar gradient dominate the tails at certain scales
that correspond to high \emph{Reynolds number} regimes, resulting in
heavy-tailed distributions. These outlier large gradients of the
passive scalar, prevailing at the inertial scales (intermittency), can
effectively be studied by means of \emph{instanton calculus} due to
their low probabilities, which forms the main contribution of this
work.

As we will line out below, the instanton
formalism~\cite{grafke-grauer-schindel:2015}, and its more rigorous
cousin, large deviation theory~\cite{freidlin-wentzell:2012}, rely on
the fact that in stochastic systems rare events often occur in a
rather predictable way: While common events usually have a multitude
of possible histories, outlier events must rely on a very precise
interplay of physical mechanisms and forcing realizations, leading to
a prototypical system trajectory for the desired rare event. At its
core lies the estimation of a stochastic (path-)integral by a
saddle-point approximation, or equivalently by a (functional) Laplace
method, that computes the most likely trajectory, called the
\emph{instanton}, as well as its probability, as the solution of a
large optimization problem. Instanton calculus has been successfully
applied to many stochastic systems, including in fluid
dynamics~\cite{balkovsky-falkovich-kolokolov-etal:1997,
  grafke-grauer-schaefer-etal:2014, grafke-grauer-schaefer:2015,
  laurie-bouchet:2015} and
waves~\cite{dematteis-grafke-onorato-etal:2019,
  tang-vanden-eijnden-stadler:2020}. These principles will be applied
in this paper to analyze outlier events in passive scalar
turbulence. More specifically, we will investigate extreme gradients
of $\theta$ for PS-RFD models via the instanton formalism to find the
most likely realization leading to outlier events, and compare the
probability scaling predicted by the instanton to the observed
heavy-tailed distribution of Monte Carlo simulations. This
demonstrates how the instanton gives us direct access to the tail
scaling of passive scalar turbulence.

This paper is structured as follows: Section~\ref{Sec:RFD} provides a
brief overview of the RFD models of the flow velocity gradient and the
passive scalar gradient. Following that, in section~\ref{Sec:
  reduced_section}, we introduce a reduced version, based on axial and
reflection symmetry considerations that are obeyed statistically by
the system. We will investigate the limitations of these symmetry
assumptions and the symmetry breaking of large strain events in
section \ref{Sec: Numerical results of RFD}. Section
\ref{Sec:Instanton_formalism} is devoted to the instanton formalism as
applied to the PS-RFD system, including its action/rate function and a
system of instanton equations that solve the optimization
problem. Section~\ref{Sec: Numerical results} then analyses
heavy-tailed PDFs of the passive scalar gradient. Such heavy-tailed
distributions, associated with non-convex rate-functions, pose a
particular difficulty for the application of sample path large
deviations; thus, we apply in section~\ref{Sec:Result_Extreme_PS} a
revised formalism based on nonlinear convexification of extreme event
instantons~\cite{alqahtani-grafke:2021}. Finally, we conclude in
section~\ref{Sec:Conclusion}.

\section{The recent fluid deformation models}
\label{Sec:RFD}

In this section, we briefly recall the recent fluid deformation
model~\cite{chevillard-meneveau:2006} and its extension to the
dynamics of passive scalar gradients~\cite{gonzalez:2009}.

\subsection{Lagrangian velocity gradient in the recent fluid deformation model}

The Lagrangian time evolution of the velocity gradient tensor $\A$ is
obtained by taking the gradient of the NSE (\ref{eq:NSE}):
\begin{equation}
  \frac {dA_{ij}}{dt} = - A_{in} A_{nj}- \frac {\partial^2 p}{\partial x_i \partial x_j}
  + \nu \frac{\partial^2 A_{ij}}{\partial x_n \partial x_n}, \label{eq:LagA}
\end{equation}
where $d / dt = \p / \p t + u_k \, \p / \p x_k$ stands for the
material derivative.  Due to the incompressibility of the flow, $\A$
must be traceless, $\Tr (\A) = 0$. As previously stated, equation
(\ref{eq:LagA}) is not closed in terms of $\A$ at position $x$ and
time $t$ because the anisotropic part of the pressure Hessian is
highly non-local and the Laplacian of $\A$ in the viscous term is not
easily expressed in terms of $\A$.

The RFD closure models these unclosed terms based on the hypotheses
detailed in appendix \ref{Appendix1}. The RFD dynamics of the
deformation that the Lagrangian particle undergoes along the flow,
(\ref{eq:LagA}), is:
\begin{equation}
\label{eq:RFD}
\frac{d \A}{dt} = - \A^2 +  \frac{\Tr( \A^2) }{\Tr(\C^{-1})} \, \C^{-1} 
- \frac{\Tr(\C^{-1})}{3T} \A  + \sqrt{\varepsilon} \, \mathbb{W} \, ,
\end{equation}
where $\C$ approximates the Cauchy-Green tensor, and a tensorial
stochastic force $\mathbb{W}\left(t\right)$ has been introduced to
produce stationary statistics. Its strength is determined by a
parameter $\varepsilon$. It is correlated as
\begin{equation*}
  {\EE \left[ dW_{ij}(t) dW_{kl}(t)\right] = G_{ijkl} \, dt,}
\end{equation*}
where the fourth order tensor, $${G_{ijkl} = 2 \delta_{ik} \delta_{jl} 
- \frac{1}{2} \delta_{il} \delta_{jk}- \frac{1}{2} \delta_{ij} \delta_{kl}},$$ is 
consistent with both the isotropy assumption and the fluid's incompressibility.

For~(\ref{eq:RFD}), the Reynolds number is defined by the ratio ${\Re \sim
  (\tau/T)^{-2}}$ of the two time scales present in the system: the
\emph{decorrelation time} $\tau$, which is assumed to be the
Kolmogorov time scale, and the \emph{integral time scale}
$T$. Suitably, we shall rephrase equation (\ref{eq:RFD}) in a
non-dimensionalized form,
\begin{equation}
\label{eq:non_dim_RFD}
\frac{d \bar{\A}}{d\bar t} = - \bar{\A}^2 +  \frac{\Tr( \bar{\A}^2) }{\Tr(\bar{\C}^{-1})} \, \bar\C^{-1} 
- \frac{\Tr(\bar{\C}^{-1})}{3} \bar\A  + \sqrt{\bar\varepsilon} \, \bar{\mathbb{W}} \, ,
\end{equation}
where the dimensionless variables are defined according to,
\begin{align}
\label{eq:non_dim}
\bar{t} = \frac{t}{T}\,, \quad \bar{\A} &= T \A \, , \quad \bar{\tau} = \frac{\tau}{T} \,, \quad \bar \C  = \C \, , \nonumber \\
\bar \varepsilon &= T^3 \varepsilon \, , \quad \bar{\mathbb{W}} = \sqrt{T}\, \mathbb{W} \, .
\end{align}
We highlight that, while the effect of the Reynolds number is not
directly stated in (\ref{eq:non_dim_RFD}), it resides in the
Cauchy-Green tensor $\C$, which in dimensionless variables can be
recast as
\begin{equation}
\label{CG_dim}
\bar \C = \exp \left( \frac{\bar \A}{\sqrt{\Re}}\right) \exp\left( \frac{ \bar \A^{\text{T}} }{\sqrt{\Re}}\right)
\end{equation}
(compare to its dimensional version  (\ref{eq:C_tensor})).

\subsection{Passive scalar turbulence in the recent fluid deformation model}

In a similar manner, taking the gradient of the PSE (\ref{eq:PS})
yields
\begin{equation}
\frac{d \psi_i }{dt} = - A_{ji} \, \psi_j + \kappa \, \frac{\partial^2 \psi_i}{\partial x_j \partial x_j}; 
 \quad  \psi = \nabla \theta \in \mathbb{R}^3.
\label{eq:Lag_psi}
\end{equation}
Following the same rationale of the previous section, 
the PS-RFD is derived from closing the diffusive Laplacian with the help of 
the short-time Cauchy-Green tensor and a \emph{diffusive integral time} scale $T_\theta$, yielding  
\begin{equation}
\label{eq:PSRFD}
\frac{d \psi}{dt}  = - \A^{\text{T}} \, \psi  - \frac{\Tr (\C^{-1})}{3T_\theta} \psi + \sqrt{\varepsilon} \, F \,, 
\end{equation}
where $F$ denotes a random force that is white in time with amplitude
$\varepsilon$, whose correlation reads ${\EE \left( dF_i(t) \, dF_j(t)
  \right) = \delta_{ij} \, dt}$. Hereafter, we assume that the noise
strength is the same in both stochastic equations (\ref{eq:RFD}) and
(\ref{eq:PSRFD}). \cite{gonzalez:2009} investigates the statistical
characteristics of the kinematics of the RFD passive gradient, whereas
\cite{hater-homann-grauer:2011} compares the PDFs from
(\ref{eq:PSRFD}) and the DNS, revealing the presence of heavy tails.

In terms of dimensionless variables \eqref{eq:non_dim}, the PS-RFD becomes,
\begin{equation}
\label{eq:non_dim_PSRFD}
\frac{d \bar \psi}{d \bar t} = -\bar \A^{\text{T}} \bar \psi - \frac{\Tr (\bar \C^{-1})}{3 \bar{T}_{\theta}}  \bar \psi 
+ \sqrt{\bar \varepsilon} \, \bar{F} \, ,
\end{equation} 
where $\bar \psi = T \psi$, $\bar F = \sqrt{T} F$ are introduced
as the dimensionless passive scalar gradient and random forcing,
respectively, $\bar{T}_{\theta}$ is the dimensionless diffusive
constant and $\bar \C$ is provided by (\ref{CG_dim}).  It is tempting
to identify the dimensionless time scale with the Schmidt $Sc$ number
as it measures the ratio $\nu/\kappa$. However, see Appendix
\ref{Appendix1}, $Sc = \nu/\kappa \approx (T_{\theta}/T)((\partial
X)^2/(\partial X_\theta)^2)$. The assumption made by the model
considers $\partial X = \partial X_\theta$, that is, the smallest
scales of turbulence are of the same order of the smallest scales of
the diffusive process. It is known from the phenomenology of
turbulence that these length scales are of the same order for $Sc$
near unity. As a result, the PS-RFD is limited to $Sc$ close to unity
\cite{gonzalez:2009}.  The role of $\bar{T}_{\theta}$ and $\Re$ in the
development of extreme events shall be discussed in section \ref{Sec:
  Numerical results}. Subsequently, we will be working with the
dimensionless RFD and PS-RFD with the bar suppressed for notational
clarity.

\section{Reduced RFD and passive scalar RFD models}
\label{Sec: reduced_section}

Conditioning on large strain values in the RFD model, and similarly on
large passive scalar gradients in the PS-RFD system reveals a
statistical tendency to respect axial and reflective symmetries around
the axis prescribed by the dominant strain. This has been observed
before for the RFD model~\cite{grigorio-bouchet-pereira-etal:2017},
and for PS-RFD~\cite{grigorio:2020}, leading to a simplification of
both RFD and PS-RFD models. This motivates us here to discuss some
details of this dimensional reduction, in particular how spontaneous
symmetry breaking at large strain values leads to a failure of the
symmetry-based reduction.

\subsection{Dimensional reduction of the RFD model}

The RFD model~(\ref{eq:RFD}) describes the evolution of a $3\times3$
matrix $\A$, but in fact has only 5 independent variables: This is
easily understood following the standard decomposition of the velocity
gradient into symmetric and anti-symmetric parts, namely, $A_{ij} =
S_{ij} + \Omega_{ij}$ where $S_{ij} = (A_{ij}+A_{ji})/2$ and
$\Omega_{ij} = (A_{ij}-A_{ji})/2$ represent the rate of strain and
rate of rotation tensors, respectively. By diagonalizing $S_{ij}$,
only three of the six variables in $S_{ij}$ remain.  The
interpretation is that after diagonalization, the coordinate
system is aligned with the principal axis of strain, from which only
two are independent due to $\Tr(\A) = \Tr(S) = 0$. The rotation
matrix's three variables represent the rate of rotation with respect
to each principal axis. Explicitly,
\begin{equation}
\A = \begin{bmatrix}
 a & 0 & 0 \\
 0 & b & 0 \\
 0 & 0 & c
\end{bmatrix} + \frac{1}{2}\begin{bmatrix}
0 & -\omega_c & \omega_b \\
\omega_c & 0 & -\omega_a \\
- \omega_b & \omega_a & 0
\end{bmatrix} \,,
\end{equation}
with $a$, $b$ and $c=-(a+b)$ are the three rates of strain, and
$\omega_a$, $\omega_b$ and $\omega_c$ are the projections of the
vorticity $\omega_i = \epsilon_{ijk} \Omega_{kj}$ along the principal
axes.

Consider the case of conditioning on a large value for the first
longitudinal component of the velocity gradient, \emph{e.g.},
$A_{11}(t_f)$ takes a value $a$. It is clear that
\begin{equation}
\label{eq:A11}
A_{11} = \left(\Lambda(\alpha) \, \A \, \Lambda^{\text{T}}(\alpha)\right)_{11} \, ,
\end{equation} 
where $\Lambda(\alpha)$ is the rotation matrix with respect to $x_1$ axis, namely, 
\begin{equation}
\Lambda(\alpha) = \begin{bmatrix}
1 & 0 & 0 \\
0 & \cos\alpha & \sin \alpha \\
0 & -\sin \alpha & \cos \alpha
\end{bmatrix} \, .
\end{equation}
Equation (\ref{eq:A11}) simply means that many different
configurations of $\A$ lead to the same $A_{11}$, namely those
obtained by rotating about the $x_1$ axis, which is a manifestation
of the axial symmetry.  Indeed, by arguments of isotropy, the
probability obeys $P(\A) = P(\Lambda(\alpha)\A
\Lambda^{\text{T}}(\alpha))$.

We can, in addition, demand that $\A$ \emph{itself} is
axisymmetric. This corresponds to a situation where we assume that
only the $x_1$-component of the strain is relevant, and we are free to
ignore the others. In this case, the number of degrees of freedom can
be reduced even more. Let an infinitesimal rotation about the $x_1$
axis given by $\Lambda_{ij} = \delta_{ij} + \alpha \ \epsilon_{1ij} +
\mathcal{O}(\alpha^2)$. After this transformation, the velocity
gradient reads,
\begin{align}
\label{eq:inf_rot2}
A'_{ij} = A_{ij} + \alpha \ (\epsilon_{1ik}A_{kj}+\epsilon_{1jl}A_{il}) \, .
\end{align}
With the hypothesis that $\A$ is invariant under rotations with respect to $x_1$, that is, $A'_{ij} = A_{ij}$, 
it can be shown that $\A$ takes the form
\begin{equation}
\A = \begin{bmatrix}
a & 0 & 0\\
0 & -a/2 & -\omega_a/2 \\
0 &\omega_a/2 & -a/2
\end{bmatrix} \, .
\end{equation} 
As a result, the number of degrees of freedom was reduced from 5 to
2. One of them is related to the rate of strain, $a$, and the other is
related to the vorticity. By invoking the reflection transformation over
the $x_2$-$x_3$-plane (i.e., ${x_1 \rightarrow -x_1}$) and admitting
that $\A$ respects this symmetry as well, we have that $\omega_a =
-\omega_a = 0$, and only one degree of freedom remains.

In summary, diagonalizing the rate of strain tensor reduces the
degrees of freedom from nine to five. Furthermore, assuming invariance
of rotation about one the principal axis of strain (axial symmetry)
implies that the vorticity lines up with the principal axis, so that a
single component of the vorticity remains, decreasing the number of
independent variables by two.  Additionally, the same axial symmetry
demands the two rates of strain to be the same, which implies two
degrees of freedom left. Finally, the assumption that the velocity
gradient respects reflection symmetry requires a zero vorticity;
otherwise, the symmetry would be broken. As a result, only a single
degree of freedom is left, corresponding to the axial rate of strain.

Based on these arguments, we can devise a simplified stochastic model
that accounts for the same statistics of the longitudinal component of
the RFD model (\ref{eq:non_dim_RFD}), which we call \emph{reduced} RFD
\cite{grigorio-bouchet-pereira-etal:2017}, given by
\begin{equation}
\label{eq:1dRFD}
\frac{da}{dt} = v(a) + \sqrt{\varepsilon} \, \eta, 
\end{equation}
where $a$ corresponds to $A_{11}$ and 
\begin{equation}
\label{eq:va}
v(a) = -a^2 + \frac 3 2 a^2 \frac{e^{-\frac{2a}{\sqrt{\Re}}}}{{e^{-\frac{2a}{\sqrt{\Re}}}}+ 2 e^{\frac{a}{\sqrt{\Re}}}} 
- \frac{a}{3} (e^{-\frac{2a}{\sqrt{\Re}}}+ 2 e^{\frac{a}{\sqrt{\Re}}}) \, .
\end{equation}
The noise term $\eta(t)$ is a zero mean white scalar random variable.
 
One may ask whether the assumption of invariance under rotation of
$\A$ is always valid. The answer is no.  As it will be discussed in
section \ref{Sec: Numerical results of RFD}, there is a critical $\Re$
above which the velocity gradient $\A$ fails to share same symmetry of
the probability, and the system undergoes a spontaneous symmetry
breaking. Hence, the dimensional reduction is no longer
possible. Crucially, this critical $\Re$ coincides with similar
limitations of the original RFD model \cite{chevillard-meneveau:2006}.

\subsection{Numerical results for symmetry breaking of the RFD model}
\label{Sec: Numerical results of RFD}

\begin{figure*}
	\begin{center}
	  \includegraphics[width=\linewidth]{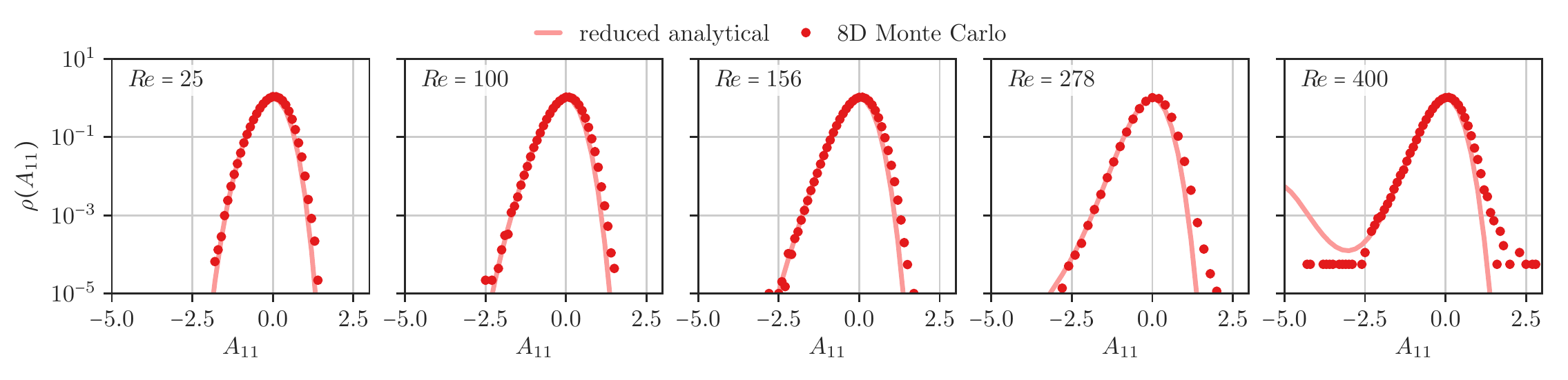}
	\end{center}
	\caption{PDFs of the $A_{11}$ component of the velocity gradient
      tensor $A_{ij} = \partial u_i / \partial x_j$, for a range of
      Reynolds numbers. The red dots show a histogram of a DNS for the
      full RFD model (\ref{eq:non_dim_RFD}), compared against the
      analytical prediction of the reduced RFD system~(\ref{eq:1dRFD})
      (solid line). It shows the emergence of another fixed point of the
      1D reduced system at $A_{11} = -3.01$ for $\Re = 400$, which is
      an artifact of the model reduction.
	\label{fig:A2Dvs11D_Plot}}
\end{figure*}

\begin{figure}
	\begin{center}
		\includegraphics[width=350pt]{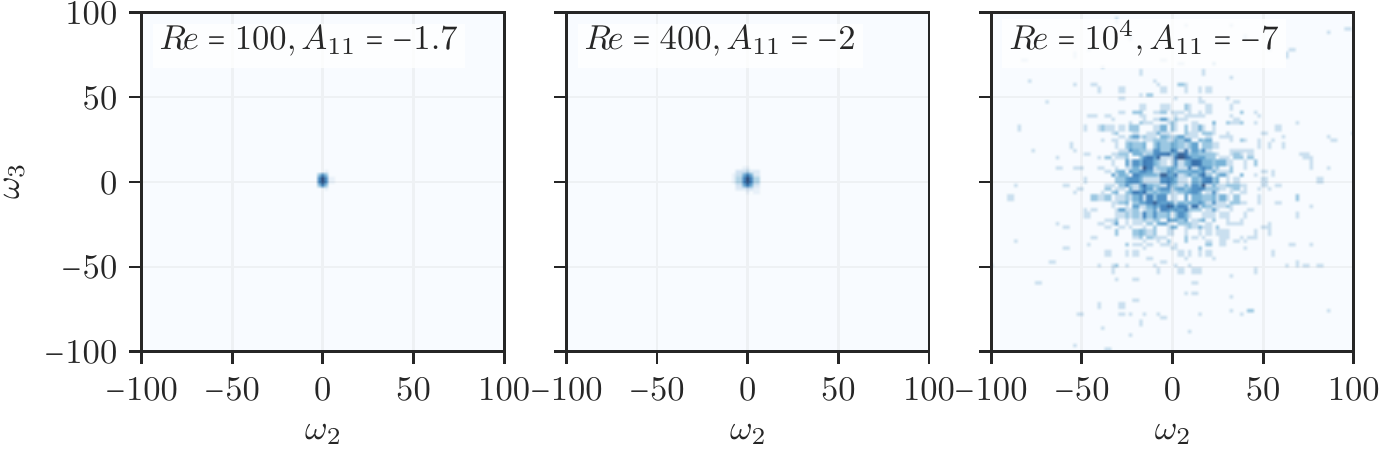}
	\end{center}
	\caption{The joint PDF $\rho(\omega_2, \omega_3)$ of the
      perpendicular components of the vorticity, conditioned on
      relatively large negative strain values $A_{11}$, for different
      values of $\Re$. As $\Re$ increases, the originally very small
      variances around the most likely configuration
      $(\omega_2,\omega_3)=(0,0)$ grows in variance. At high $\Re$, a
      zero perpendicular vorticity is no longer predominant.  Instead,
      the most likely vorticity clusters on a ring, indicating a
      spontaneous symmetry breaking of the vorticity conditioned on
      large negative strain.
	\label{fig:Ring_Plot}}
\end{figure}

Here we give evidence for the validity of the dimensionally reduced
model~(\ref{eq:1dRFD}) for moderate $\Re$, and the eventual symmetry
breaking of the full 8D model~(\ref{eq:non_dim_RFD}). We recall that
for the reduced RFD, an analytical PDF can be easily found by solving
the corresponding Fokker-Planck equation
\cite{grigorio-bouchet-pereira-etal:2017}. Shown in
figure~\ref{fig:A2Dvs11D_Plot} are the PDFs obtained via Monte Carlo
(MC) simulations (red dots), in the range $\Re \in [25,400]$, against
the analytical PDFs of the reduced model (solid red lines). For the
lowest $\Re$ values up to $\Re = 156$, there is a reasonable agreement
between the full 8D-RFD and the reduced 1D-RFD. For higher $\Re$, a
disagreement is seen in the right tail; note, though, that positive
strain values are irrelevant for the development of large velocity
gradients, as will be shown later.  As $\Re$ is increased further to
$\Re = 400$ ($\tau/T = 0.05$), at the very right of
figure~\ref{fig:A2Dvs11D_Plot}, the disagreement becomes more
pronounced, including on the far left tail. Here, the 1D-RFD predicts
a bimodal PDF with a new local minimum located at $A_{11} = -3.01$. By
contrast, this bimodality is not observed in the 8D-RFD. The emergence
of this bimodal profile remains for larger values of Reynolds number.
Roughly $\Re=400$ establishes the upper limit where the dimensional
reduction can sensibly be applied. 
  
The discrepancy between 8D-RFD and 1D-RFD demonstrates that for $\Re
\gtrsim 400$ the hypothesis of symmetries (axial and reflection)
outlined in the previous section do not hold. As a consequence, other
components of the velocity gradient start to play a role in the
dynamics and may not be neglected. However, it remains true that the
equation itself, and thus also the PDF, remains invariant under
rotations and reflections for any value of the parameter $\Re$. It is
only individual sample trajectories that break the symmetry, while the
statistics remain symmetric. Hence, it makes sense to borrow a
terminology of condensed matter/high-energy physics, observing that
the model undergoes spontaneous symmetry breaking, since the symmetry
of the model is not realized by the individual states of the system
$\A$, even though the PDF does observe it. The fact that this indeed
happens can be shown numerically. Figure~\ref{fig:Ring_Plot} shows the
joint PDFs $\rho(\omega_2, \omega_3)$ of the perpendicular components
of the vorticity $\omega_i = \epsilon_{ijk} \Omega_{kj}$, conditioned
on relatively large negative $A_{11}$ at different $\Re$. In other
words, this shows the distribution of the vorticity vector in the
presence of extreme strain, in the plane perpendicular to the strain
axis.  For moderate $\Re = 100$ and $400$ the distribution is
concentrated around $0$, highlighting that the vorticity vector points
along the strain axis (or is altogether zero). For very large $\Re$,
though, the perpendicular vorticity components prefer to occupy a ring
away from $(0,0)$, indicating the breakdown of axisymmetry for the
individual sample. At this $\Re$, vorticity is more likely to be at an
angle against the strain axis. Note that while we do not believe that
the RFD model remains a valid description of 3D NSE turbulence in this
regime, we remark that symmetry breaking \emph{has} recently been
observed for extreme strain events in full 3D
Navier-Stokes~\cite{schorlepp-grafke-may-etal:2021}.
 
It is worth mentioning that in same range of $\Re$ where the symmetry
breaking happens, the RFD model itself becomes problematic as well, as
numerical instabilities start to appear, as reported by
\cite{chevillard-meneveau:2006}.  Here, we shall briefly explain that
by considering the high-$\Re$ limit of \eqref{eq:non_dim_RFD}. In the
limit of infinite $\Re$, the RFD model reduces to
\begin{equation}
\label{eq:high_Re_RFD}
\frac{d \A}{dt} = - \A^2 +  \frac{\Tr( \A^2) }{3}  
- \A  + \sqrt{\varepsilon} \, \mathbb{W} \, .
\end{equation}         
Apart from the stochastic forcing, this equation corresponds to the
linear damping closure proposed by
\cite{martin-dopazo-valino:1998}. It has been shown that the linear
damping is not enough to counteract the strong non-linearities of the
self-stretching and pressure Hessian terms, being subject to finite
time singularities.

\subsection{Dimensional reduction of the passive scalar RFD model}

Following the same logic, one may derive a 2-dimensional reduced model
for the PS-RFD by considering the statistics of only a single
component of the passive scalar gradient, say $\psi_1 = \nabla_1
\theta$. More specifically, assuming now that both $\A$ and $\psi$ are
invariant under rotation around the $x_1$ axis, the components
$\psi_2$ and $\psi_3$ must vanish. As a result, the reduced version of
the PS-RFD model (\ref{eq:non_dim_PSRFD}) is defined as
\cite{grigorio:2020},
\begin{equation}
\label{eq:2dPSRFD}
\frac{d\psi_1}{dt} = b(\psi_1,a) + \sqrt{\varepsilon}\,\xi, 
\end{equation}
where 
\begin{equation}
\label{eq:detPSRFD}
b(\psi_1,a) = -\psi_1 \, a - (e^{-\frac{2a}{\sqrt{\Re}}}+ 2 \, e^{\frac{a}{\sqrt{\Re}}}) \frac{\psi_1}{3 \, T_\theta} \, ,
\end{equation} 
and $\xi(t)$ is a white scalar noise that is independent of $\eta(t)$
in~(\ref{eq:1dRFD}).  The dynamics of $\psi_1(t)$ depend on the
longitudinal velocity gradient $a(t)$. Thus, equation
(\ref{eq:2dPSRFD}) has to be solved together with (\ref{eq:1dRFD}).

Being dependent on the RFD, it is clear that the dimensional
reductions for PS-RFD will fail in the same range of $\Re$ where RFD
symmetry breaks down, but is in excellent agreement for $\Re\lesssim400$.

\section{Instanton formalism and extreme events}
\label{Sec:Instanton_formalism}

In this section, we apply the instanton formalism to the PS-RFD model
described in section~\ref{Sec:RFD}. Intuitively, the instanton
formalism relies on the fact that in some limit (such as the small
noise or extreme event limits) probabilities can be efficiently
estimated through a prototypical ``placeholder'' event that observes
the same scaling as the actual probability. A probability of an event
is always a sum (or integral) over all possible ways the event can
occur, weighted by its respective probability. In the limit, this
integral can be approximated by a saddlepoint approximation or Laplace
method, giving the leading order exponential contribution. For
example, we are interested in the probability of observing events of
extreme passive scalar gradients at final time,
$P(\psi_1(t_f)>z)$. Then, the instanton formalism postulates that the
probability scales like an exponential,
\begin{equation}
  P(\psi_1(t_f)>z) \sim \exp\left(-\eps^{-1} I(z)\right)\,.
\end{equation}
The exponential scaling, given by the \emph{rate function} $I(z)$, can
be obtained by evaluating an action $S[\A,\psi]$ at the
\emph{instanton} $(\A^*, \psi^*)$,
\begin{equation}
  \label{eq:I_z}
  I(z) = S[\A^*,\psi^*] = \inf_{\psi_1(t_f)>z} S[\A,\psi]\,,
\end{equation}
where the instanton is the minimizer of the action. We will derive the
action for the PS-RFD model in section~\ref{Sec:InstantonEqs}. In our
setup, the instanton formalism is equivalent to sample path large
deviation theory~\cite{varadhan:1966, dembo-zeitouni:2010,
  freidlin-wentzell:2012}.

\subsection{Related works}

The action functional for the RFD model has first been determined in
\cite{moriconi-pereira-grigorio:2014}.  The instanton equations were
linearized in this reference to derive an approximate analytical
solution, with additional consideration of the fluctuations around the
linearized instanton. As a result, to leading order in the
perturbative expansion, the fluctuations yield an effective action
with renormalized noise. That is, to first order, the fluctuations
around the instanton can be taken into account by renormalizing the
noise correlator. This approach was used to evaluate the PDFs of the
velocity gradient and the joint PDF of the $R$ and $Q$ invariants.

By contrast, \cite{grigorio-bouchet-pereira-etal:2017} determines the
instanton numerically by solving the corresponding highly non-linear
RFD Hamilton's equations with the Chernykh-Stepanov algorithm
\cite{chernykh-stepanov:2001}.  Further, following the perturbation
techniques outlined in \cite{moriconi-pereira-grigorio:2014}, a
detailed analytical treatment of the RFD closure has been given by
\cite{apolinario-moriconi-pereira:2019}, providing a hierarchical
classification of several Feynman diagrams. In addition to the noise
renormalization, \cite{apolinario-moriconi-pereira:2019} also computes
the propagator renormalization derived from linear instanton
approximation. The resulting PDFs are compared with the ones from
\cite{grigorio-bouchet-pereira-etal:2017} with good agreement.

More recently, \cite{grigorio:2020} applies instanton arguments also
to the PS-RFD model, proposing a parametric form of the Hamilton's
equation. Aside from that, a perturbation expansion has been carried
out along the lines of \cite{moriconi-pereira-grigorio:2014,
  apolinario-moriconi-pereira:2019} to account for instanton path
fluctuations.

Putting these results into perspective, all are capable of obtaining
only mild non-Gaussian PDFs, that is, they work for a restricted range
of $\tau$, namely, $\tau/T \geq 0.1$ ($\Re \lesssim 100$). As $\Re$
increases, and intermittency starts to play a role, the probability
distributions develop heavy tails. Consequently, the corresponding
rate function ceases to be convex, which prevents naive instanton
approaches based on the G\"artner-Ellis theorem to remain
well-posed. To overcome this, and apply the instanton formalism to
more turbulent flows, here we introduce a nonlinear convexification to
treat the heavy-tailed distribution, as discussed in section~\ref{Sec:
  Numerical results}.
  
\subsection{The action and instanton equations for the PS-RFD dynamics}
\label{Sec:InstantonEqs}

In accordance with \cite{grigorio:2020}, the PS-RFD action reads
\cite{martin-siggia-rose:1973,janssen:1976,dominicis:1976} ,
\begin{align}
\label{eq:action_PSRFD}
S[\Pe, \A, \Pi,  \psi] =  \int_{t_i}^{t_f} \text{d}t \, &\left[ \text{Tr} \left( \Pe^{\text{T}} (\dot \A - \V(\A)) \right)- \frac{1}{2} P _{ij} G_{ijkl} P_{kl} \right. \nn \\
& \left. + \Pi^{\text{T}} (\dot \psi - M (\psi,\A)) - \frac{1}{2} \Pi^{\text{T}} \Pi \right] \, ,
\end{align}
where 
\begin{equation}
\label{eq:det_theta}
M(\psi,\A) = - \A^{\text{T}} \psi  - \frac{\Tr (\C^{-1})}{3 T_\theta} \psi \, ,
\end{equation}
and
\begin{equation}
\label{eq:detA}
\V(\A) = - \A^2 +  \frac{\C^{-1} \Tr( \A^2) }{\Tr(\C^{-1})} - \frac{\Tr(\C^{-1})}{3} \A \, ,
\end{equation}
stand for the drift terms of equations (\ref{eq:non_dim_PSRFD}) and
(\ref{eq:non_dim_RFD}), respectively, and $\Pi \in \RR^3$ ($\Pe \in
\RR^{3\times 3}$) is the conjugated momentum of $\psi$ ($\A$), closely
related to the auxiliary variables of the
Martin-Siggia-Rose-Janssen-de\,Dominicis formalism
\cite{martin-siggia-rose:1973,janssen:1976,dominicis:1976}.

The minimum of the action functional \eqref{eq:action_PSRFD} is
achieved by the solutions of the following corresponding instanton
equations of the fields $\A, \psi$:
\begin{equation}
  \begin{split}
   \frac{\delta S}{\delta P_{ij}}  & = 0, \ \Rightarrow \  \dot{A}_{ij} = V \left(\A\right)_{ij} +  G_{ijkl} \, P_{kl}, \\
    \frac{\delta S}{\delta A_{ij}} & = 0, \ \Rightarrow  \ \dot{P}_{ij}   = - P_{kl} \,\, \nabla_{A_{ij}} V\left(\A\right)_{kl} - \Pi_{k} \,\, \nabla_{A_{ij}} M\left(\psi, \A \right)_{k}, \\
    \frac{\delta S}{\delta \Pi_{k}} & = 0, \ \Rightarrow \ \dot{\psi}_{k}  = M\left( \psi, \A\right)_{k} +  \Pi_{k}, \\
    \frac{\delta S}{\delta \psi_{k}} & = 0, \ \Rightarrow \ \dot{\Pi}_{k}  = - \Pi_{n} \,\, \nabla_{\psi_{k}} M\left(\psi, \A\right)_{n}\,, \label{eq:Inst_eqs}
  \end{split}
\end{equation}
for $t\in[t_i, t_f]$. The full formulas for these gradients are
derived in appendix \ref{Appendix2}, where we expand the drifts up to
second order~\cite{moriconi-pereira-grigorio:2014}.

These four coupled equations (\ref{eq:Inst_eqs}) are solved
simultaneously using the \emph{Chernykh-Stepanov} (C-S) scheme
\cite{chernykh-stepanov:2001, grafke-grauer-schaefer:2013}, which
corresponds effectively to a gradient descent of the constrained
optimization problem~\cite{grafke-vanden-eijnden:2019}. The boundary
conditions of (\ref{eq:Inst_eqs}) are specified by the choice of
observable. Here, we are looking for events where one component
$\psi_j$ of the passive scalar gradient exceeds a threshold $z$, which
leads to
\begin{equation}
\label{eq:cond}
\A \left( t_i\right) = \textbf{0}, \ \ \psi\left( t_i\right) = 0, \ \ \Pe\left( t_f \right) = \textbf{0}, \ \
\Pi_{j} \left( t_f \right) = \lambda \, \nabla F\left( \psi_{j} \left( t_f \right) \right),
\end{equation} 
where the initial values of the fields are their stable equilibrium
points, the origin.  The final time constraint on the gradient of
passive scalar to attain $z = \psi_{j} \left(t_f \right)$ is
implemented in (\ref{eq:cond}) through a Lagrange multiplier $\lambda
\in \RR$, \cite{rindler:2018}. The function $F: \RR \to \RR$ is a
nonlinear reparametrization to ensure there is a unique $\lambda$ for
every large passive scalar gradients of interest
\cite{alqahtani-grafke:2021}.

\subsection{Instantons for the reduced PS-RFD dynamics}
\label{Sec: reduced_inst}

The full instanton equations (\ref{eq:Inst_eqs}) correspond to the
system (\ref{eq:non_dim_RFD}),~(\ref{eq:non_dim_PSRFD}). However, when
a final time constraint is imposed on a component of the passive
scalar, such as $F\left( \psi_{1} \left(t_f \right) \right)$
(\ref{eq:cond}), it exhibits symmetric behavior (with respect to
axial and reflective symmetries) that reduce this 11-variables system
to one with only two leading variables, $\psi_{1} \left(t \right)$ and
$a \left(t \right)$. The same reduction applies to conditioning on
other components of $\psi$. As discussed in section~\ref{Sec:
  reduced_section}, this reduction is valid for ${\Re \lesssim 400}$.

For the reduced model~(\ref{eq:1dRFD}),~(\ref{eq:2dPSRFD}), the resulting 2D instanton
equations are
\begin{equation} 
\begin{split}
& \dot{a}  = v \left( a \right) +  \, p, \\
& \dot{p}  = - p \,\, \frac{\partial v\left(a \right)}{\partial a } 
- q \,\, \frac{\partial b\left( \psi_{1} , a \right)}{\partial a } , \\
& \dot{\psi}_{1}  = b\left( \psi_{1}, a \right) +  q, \\
& \dot{q}  = - q \,\, \ \frac{\partial b\left( \psi_{1}, a  \right)}{\partial \psi_{1}} , 
\label{eq:Inst_eqs_reduced}
\end{split}  
\end{equation} 
where $p\left(t\right) = P_{11} \left(t\right)$ and $q\left(t\right) =
\Pi_{1} \left(t\right)$.  The drifts $v \left(a \right)$ and $b\left(
\psi_{1}, a \right)$ are derived in the reduced models' section,~
\ref{Sec: reduced_section}, namely equations (\ref{eq:va},
\ref{eq:detPSRFD}).
\begin{figure}
	\begin{center}
	  \includegraphics[width=246pt]{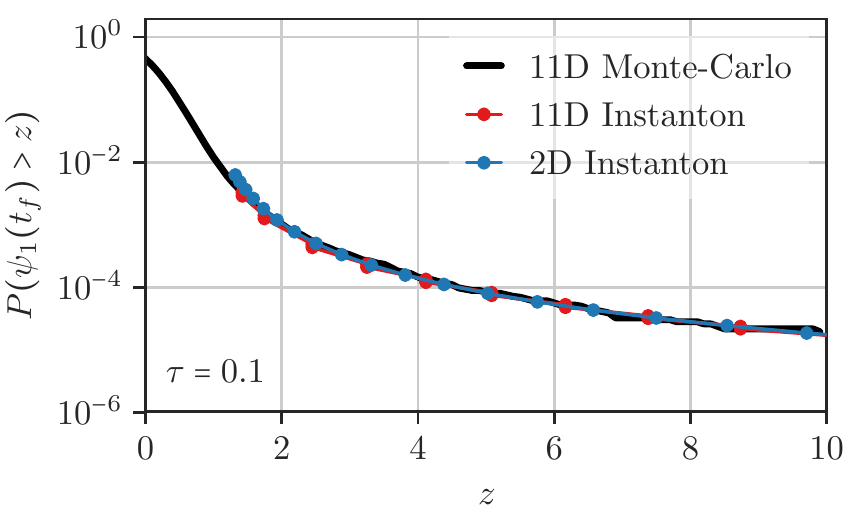}
	\end{center}
	\caption{The complementary cumulative distribution functions
	 of passive scalar gradients $P(\psi_{1}(t_f)>z)$, (solid line) compared
      to the outcomes of both 11D and 2D instantons results (red and
      blue, respectively), with $\Re = 100$ and $T_\theta = 1$.  Both
      the full and the reduced instantons agree with MC simulations of
      the full system.
		\label{fig:11Dinst_vs_2Dinst_tau0_1}}
\end{figure}
 
The difference between equations (\ref{eq:Inst_eqs}) and
(\ref{eq:Inst_eqs_reduced}) is that the latter is more computationally
efficient than the former due to the significant reduction of its
dimensions, and we will use it in the following to estimate the tail
probabilities of the passive scalar
gradient. Figure~\ref{fig:11Dinst_vs_2Dinst_tau0_1} demonstrates
numerically that this simplification is indeed justified for the
instanton, as the predicted probabilities $P(\psi_1(t_f)>z)$ of
exceeding a passive scalar gradient $z$ at final time $t_f$ is in
excellent agreement between MC sampling of the full RFD model, and the
instanton estimates of both the full and the reduced models.

\section{Extreme gradient of the passive scalar}
\label{Sec: Numerical results}

In this section we provide both analytical and numerical results for
extreme passive scalar gradients in the PS-RFD. Starting from the
model equations \eqref{CG_dim} and \eqref{eq:non_dim_PSRFD}, which we
rewrite here for convenience,
\begin{equation}
\label{eq:non_dim_PSRFD_2}
\frac{d \psi}{dt}  = - \A^{\text{T}} \, \psi  - \frac{\Tr (\C^{-1})}{3T_\theta} \psi + \sqrt{\varepsilon} \, F \,;  \quad
 \C = e^{\frac{ \A}{\sqrt{\Re}}} e^{\frac{ \A^{\text{T}}}{\sqrt{\Re}}} \,,
\end{equation} 
recall that the first term on the right hand side of this stochastic
equation accounts for the advection, whereas the second term describes
the effect of diffusion. We shall discuss the role played by
parameters $T_\theta$ and $\Re$. In the limit of high $\Re$, the
Cauchy-Green tensor $\C$ can be expanded to order
$\mathcal{O}(\Re^{-1})$,
\begin{align}
\label{eq:TrC}
\Tr(\C^{-1}) &= 3 + \frac{1}{2\Re}\Tr \left( \A^2 +     \A^{2 \text{T}} + 2 \A^{\text{T}} \A \right) \, \nonumber \\
& = 3 + \frac{2}{\Re}\Tr \left( S^2 \right) \,,
\end{align}
where, $S = (\A+ \A^{\text{T}})/2$. Taking this into consideration equation \eqref{eq:non_dim_PSRFD_2} is rewritten as
\begin{equation}
\label{eq:approx_PSRFD}
\frac{d \psi}{dt}  = - \A^{\text{T}} \, \psi  - \frac{\psi}{T_\theta} -\frac{2\psi}{3T_\theta \,\Re} \Tr(S^2)+ \sqrt{\varepsilon} \, F \,.
\end{equation}
From \eqref{eq:approx_PSRFD}, it is clear that the second term on the
right side is a linear damping for $\psi$, acting to decrease the size
of fluctuations, with $T_\theta$ being the (dimensionless)
characteristic time. The behavior of the third term, on the other
hand, can be understood as follows: To leading order $\mathcal
O(\Re^0)$, the variance of $\Tr(S^2)$ for the RFD depends on
$\varepsilon$, with subleading correction of order $\mathcal
O(\Re^{-1})$ \citep{grigorio-bouchet-pereira-etal:2017}. Hence, we
expect that as $\Re$ increases while $T_\theta$ remains constant, the
effect of the third term decreases on average.

Our claim is that, as $T_\theta$ or $\Re$  is increased the damping effects are lowered in comparison to the 
advection term, which dominates the dynamics of the passive scalar gradient $\psi$. In turn, due to the 
minus sign accompanying the advection term, this transport term will drive an increase of a given 
component of $\psi$ as long as the eigenvalue of $\A$ along the same direction is negative, allowing 
for a growth of $\psi$ to extreme values. 

\begin{figure}
  \begin{center}
    \begin{minipage}{246pt}
      \includegraphics[width=\textwidth]{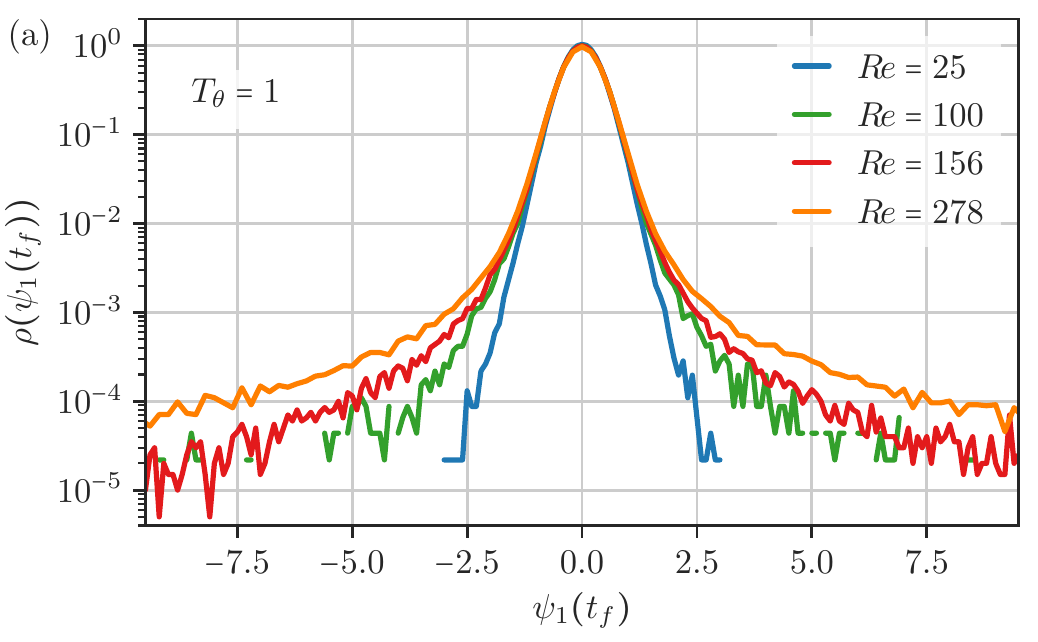}\\
    \end{minipage}
    \begin{minipage}{246pt}
      \includegraphics[width=\textwidth]{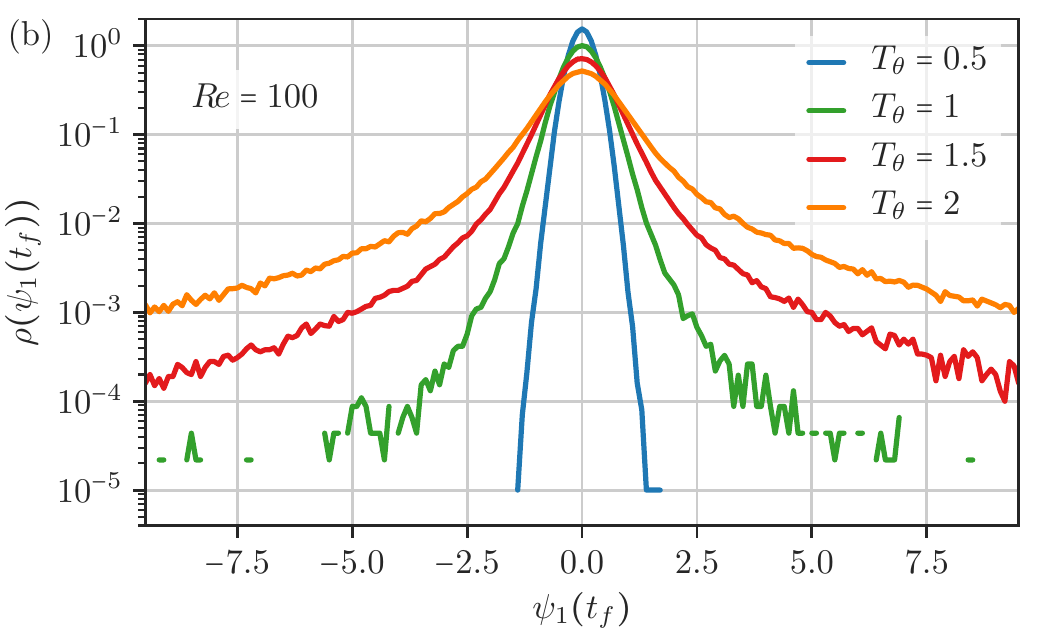}
    \end{minipage}
  \end{center}
  \caption{PDFs of the first component of passive scalar gradients,
    $\psi_1$, obtained from MC simulations of the PS-RFD system
    (\ref{eq:non_dim_RFD}),~(\ref{eq:non_dim_PSRFD}) for various
    Reynolds numbers $\Re$ and diffusive time scales
    $T_\theta$. Subfigure (a) exhibits heavier tails as $\Re$
    increases, where $T_\theta$ is set to unity, indicating
    significant turbulent mixing.  Subfigure (b) shows that increasing
    $T_\theta$ at a moderate value of Reynolds number, $\Re = 100$,
    results in heavy-tailed gradient distributions, caused by a high
    transport rate.
		   \label{fig:PDFs_11D_Grth1} }
\end{figure}

\subsection{High Reynolds number regime: Heavy tails and convexification}
\label{Sec:Re_No_Regime}

We are now equipped to investigate the probability to observe extreme
passive scalar gradients for different $\Re$ and
$T_\theta$. Figure~\ref{fig:PDFs_11D_Grth1} displays the PDFs of the
first component of passive scalar gradients, $\psi_1$, at the final
time $t_f$, for various values of $\Re$ in (a), and diffusive
timescales $T_\theta$ in (b). They are obtained by MC simulations of
the full 11D PS-RFD system (\ref{eq:non_dim_RFD},
~\ref{eq:non_dim_PSRFD}). It illustrates that indeed increasing both
$\Re$ and $T_\theta$ invokes heavy tails for the passive scalar
gradient, due to strong turbulent mixing and high transport rates. We
also remark that the fattening of the tails is more sensitive to the
diffusive time scale $T_\theta$ than to the Reynolds number. This can
be understood through equation \eqref{eq:approx_PSRFD}, where it is
evident that increasing $T_\theta$ leads to a decrease of two
suppression terms for $\psi$, compared to only one for $\Re$.

\subsection{Extreme configurations of the passive scalar gradient}
\label{Sec:Result_Extreme_PS}

\begin{figure}
	\begin{center}
		\includegraphics[width=246pt]{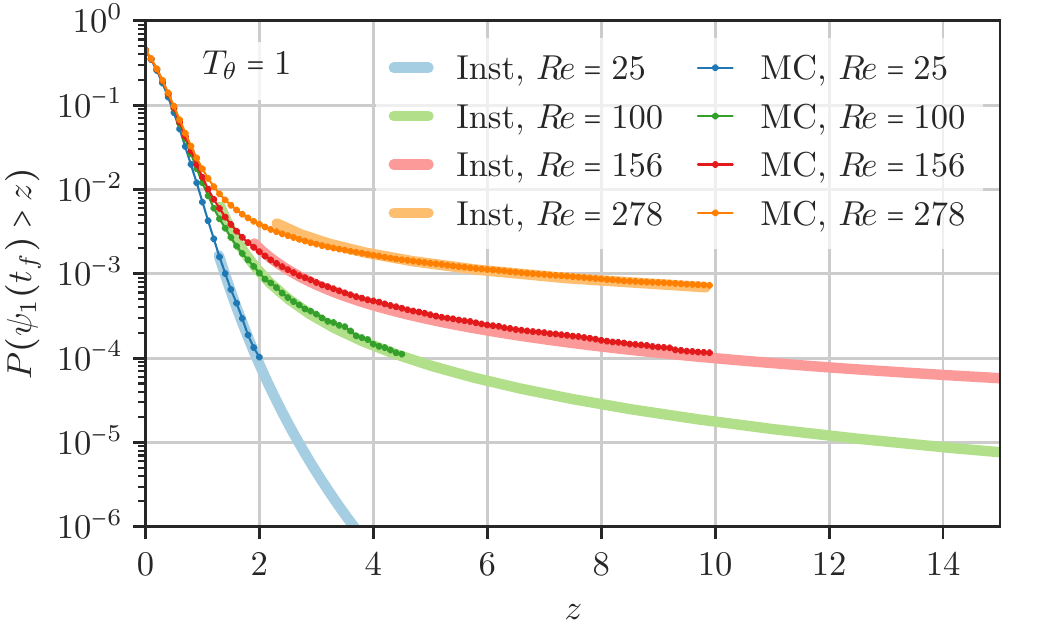}
	\end{center}
	\caption{The complementary cumulative distribution functions of
      the passive scalar, $P(\psi_{1}(t_f) >z)$.  Compared are MC
      simulations (dotted lines) against 2D instantons estimates
      (solid lines) for different values of $\Re$ ($T_\theta$ is set
      to unity). There is clear agreement between 11D MC and the 2D
      instanton estimate in particular when $\psi_1$ becomes large
      (far tails), in accordance with large deviations theory.
	\label{fig:MC11Dvs2Dinst}}
\end{figure}

The probabilities obtained from MC sampling can be directly compared
to predictions from the instanton formalism, obtained by solving the
optimization problem~(\ref{eq:I_z}). In practice, we do so by
numerically solving the instanton
equations~(\ref{eq:Inst_eqs_reduced}). This comes of a significant
performance benefit over computing the instanton for the full
model~(\ref{eq:Inst_eqs}), allowing us to compute the minimizer faster. 
For example the average speed-up factor for $\Re = 100$ and $z \in [2, 10] $ is $54$. 
The benefit of solving equations~(\ref{eq:Inst_eqs_reduced}) is 
even more significant for extreme events since this factor of 
improvement grows as $\Re$ and/or $z$ increase. 
To overcome the problem of heavy tails, we convexify
the rate function with a reparametrization of the observable according
to the scheme presented in~\cite{alqahtani-grafke:2021}. Concretely,
we choose ${F(z) = \mathrm{sign}(z) \log \log |z|}$, to be inserted as
boundary condition into~(\ref{eq:cond}). We then use the C-S
algorithm~\cite{chernykh-stepanov:2001, grafke-grauer-schaefer:2013}
to obtain the instanton fields $\A$ and $\psi$ and its respective
conjugate momenta $\Pe$ and $\Pi$. These allow us to (i) obtain the
tail scaling of the passive scalar gradient PDF, by computing the
action of the instanton, and (ii) identify the mechanism responsible
for the formation of extreme passive scalar gradient events within the
model.

\begin{figure}
	\begin{center}
		\includegraphics[width=246pt]{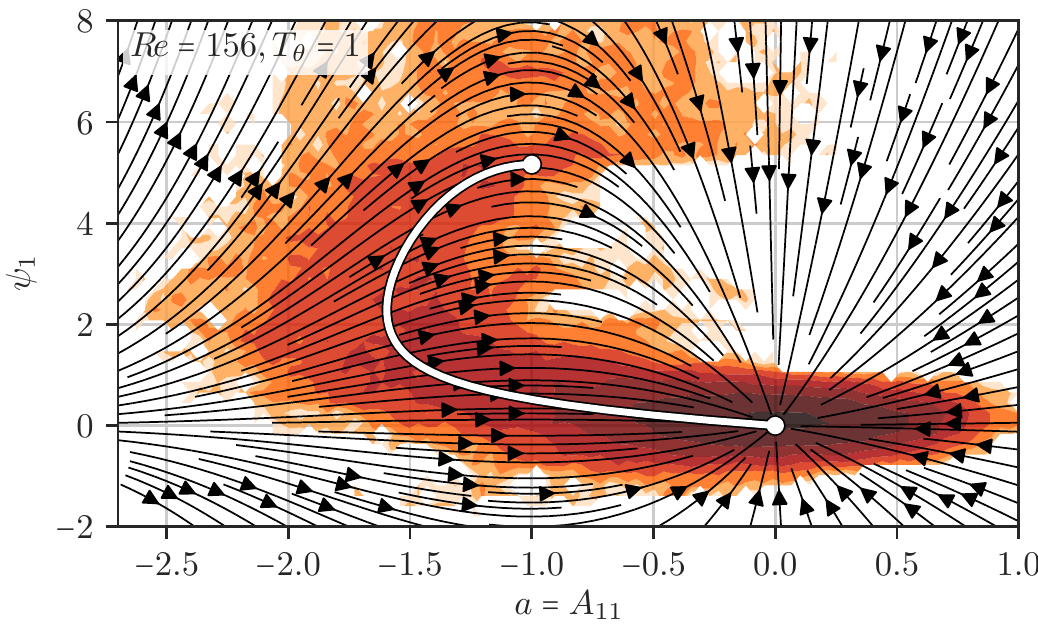}
	\end{center}
	\caption{Events of extreme passive scalar gradient for the reduced
      model in the $a$-$\psi_1$-plane. The streamlines depict the
      deterministic drift of the reduced PS-RFD model. The line shows
      the instanton for realizing a final-time event of $\psi(t_f) =
      5.16$, starting from the fixed point $(0,0)$ (white line). The
      density of trajectories of MC-simulations, conditioned on the
      same outcome, is shown as heatmap. Extreme outcomes of $\psi_1$
      are commonly achieved by first transitioning into a region of
      negative strain, in which it is much easier to excite strong
      gradients. The instanton correctly predicts this
      mechanism.\label{fig:2D_plot}}
\end{figure}

Figure~\ref{fig:MC11Dvs2Dinst} presents the logarithmic
probabilities of MC simulations of the passive scalar $\psi_{1}(t_f)$
(dotted lines) against 2D instantons results (solid lines) for
different values of $\Re$, where $T_\theta$ is set to unity. It
demonstrates an excellent agreement of the tail scaling between the
11D MC simulation and the instanton prediction, in particular when
$\psi_1$ becomes large. Note that the instanton computation allows us
to go extremely far into the tails, where MC becomes
inefficient. Figure~\ref{fig:2D_plot} depicts a set of realizations
achieving an extreme passive scalar gradient, $\psi_1=5.16$, in the
$a$-$\psi_1$ plane. Here, the shading indicates the density of
trajectories that exhibit this large passive scalar gradient at the
final time, while the solid line shows the instanton prediction for
comparison. Clearly visible is the dominant mechanism for producing
large passive scalar gradients: Fluctuations in the relevant strain
component drives the system into a region of large negative strain,
which deterministically amplifies the passive scalar gradient to large
values. Note that the dominant reactive channel is nicely predicted by
the instanton.

\section{Conclusion}
\label{Sec:Conclusion}

We investigate events of extreme passive scalar gradients in turbulent
flows by using Lagrangian turbulence models extended to handle passive
scalar advection. We demonstrate how a reduced two-dimensional model
(one component of strain and passive scalar gradient each) captures
the important mechanisms responsible for large passive scalar
gradients. Notably, the symmetries necessary to apply the reduced
model become broken for very extreme events or very large Reynolds
numbers, which we can observe by direct sampling. We remark that the
full RFD model also fails to describe fully developed Navier-Stokes
turbulence in this regime, so that the reduced model remains a helpful
simplification for our purposes.

We employ the instanton formalism to capture the scaling of very large
outlier events in the tails of the PDF of passive scalar
gradients. This most likely trajectory not only yields the correct
tail scaling, even in the fat-tailed regime, but further allows us to
investigate the mechanism responsible for the buildup of large
gradients in the reduced model.

\section*{Acknowledgments}

MA acknowledges the PhD funding received from UKSACB. TG acknowledges
the support received from the EPSRC projects EP/T011866/1 and
EP/V013319/1.

\bibliographystyle{apsrev4-1}
\bibliography{bib}

\onecolumngrid
\section{Appendix}
\appendix

\subsection{Statistical modeling of the fluid velocity gradient}
\label{Appendix1}

The hypotheses that has been used to model both of the NSE and PSE, producing the RFD closure is summarized here, as follows;
\begin{itemize}
	\item First, in Lagrangian coordinates $x\left(X, t\right)$ where $X$ is the initial position,
	the velocity is a function of time and initial position, making it feasible to visualize its history. 
	Therefore, the main hypothesis of the RFD is that, due to the uncertainty of the deformation 
	history resulting of the stochastic nature of turbulent flows, the history of the velocity 
	gradient tensor $\A(t)$ has been forgotten, thus $\A$ depends only on its \emph{recent} configuration, 
	hence the name. In other words, the velocity gradient processes have been considered 
	as Markov processes at which the future is independent of the past, given the current 
	value. 
	\item Based on the preceding hypothesis, the second assumption is that the Lagrangian 
	pressure Hessian $\p^2 p / \left( \p X_m \p X_l \right)$ is an isotropic tensor, i.e., equals 
	$a \, \delta_{ml}$, for any scalar $a$. This constant $a$ is selected in this model to be 
	one-third of its trace, that is,
	$$\frac{\p^2 p}{\p X_m  \p X_l} \approx \frac{1}{3} \frac{\p^2 p}{\p X_q  \p X_q} \delta_{ml}.$$
	As a result, using the chain rule and neglecting higher order terms, the pressure term of equation 
	(\ref{eq:LagA}) is modeled as: 
	\begin{equation}
	\label{eq:pressure_term}
	 \frac{\p^2 p}{\p x_i  \p x_j} \approx \frac{\p X_m}{\p x_i} \frac{\p X_l}{\p x_j} \frac{\p^2 p}{\p X_m  \p X_l} 
	\approx \frac{\p X_m}{\p x_i} \frac{\p X_l}{\p x_j} 
	\frac{1}{3} \frac{\p^2 p}{\p X_q  \p X_q} \delta_{ml} =
	- \frac{\Tr(\A^2 )}{\Tr(\C^{-1})} \, C_{ij}^{-1},
	 \end{equation}
	where the last equality comes from utilizing the Poisson equation to express $p$ in terms of $\A$ (i.e.~by taking 
	the divergence of the NSE (\ref{eq:NSE}) and employing the divergence free property). The tensor 
	\begin{equation}
	\label{eq:C_tensor}
	{C_{ij} = \frac{\p x_i }{\p X_m} \, \frac{\p x_j}{\p X_l} \, \delta_{ml} \approx 
		\exp\left( \tau \A \right) \exp \left( \tau \A^{\text{T}} \right) },
	\end{equation}
	is a stationary Cauchy-Green tensor, where $\tau$ is the decorrelation time scale
	after which any correlation of $\A$ is neglected, based on the main hypothesis 
	given earlier. This parameter $\tau$ is proportional to $\Re^{-1/2}$. It therefore plays an essential 
	role of the dynamics of this model, more discussion is in section~\ref{Sec:Re_No_Regime}.  
	
	\item The Lagrangian viscous Hessian of $\A$ has been treated as a classical linear damping term, i.e.,
	${\nu   \ \A / (\partial X_m  \partial X_l ) \approx - \nu \ \A / (3 T) }$ where a dimensional argument used to write 
	$ \nu / \left(\p X \right)^2 \sim T^{-1}$, and $T$ is considered to be on the order of the integral time scale 
	of the flow.  Finally, the model of the viscous term of (\ref{eq:LagA}) becomes: 
	\begin{equation}
	\label{eq:viscous_term}
	\nu \ \frac{\partial^2 \A}{\partial x_n \partial x_n} \approx \nu \
	\frac{\partial X_m}{\partial x_n} \frac{\partial X_l}{\partial x_n}   \frac{\partial^2 \A}{\partial X_m  
	\partial X_l} \approx - \frac{\Tr(\C^{-1} )}{3 T} \A.
	\end{equation}
	Notice that the diffusivity term of equation (\ref{eq:Lag_psi}) is modeled using the same assumptions of the 
	viscosity term, yielding to
	\begin{equation}
	\label{eq:diffusive_term}
	\kappa \ \frac{\partial^2 \psi}{\partial x_j \partial x_j}  \approx - \frac{\Tr (\C^{-1} )}{3 T_\theta} \psi, 
	\ \ \ \ \ \ \ \    \kappa / \left(\p X_\theta \right)^2  \sim T_{\theta}^{-1}\, ,
	\end{equation}
\end{itemize}
where $\p X_\theta$ is related to the smallest scales structures of the passive scalar. Finally, substituting the 
foregoing closed terms (\ref{eq:pressure_term},\ref{eq:viscous_term}) into the Lagrangian velocity gradient 
equation (\ref{eq:LagA}) produces the velocity gradient RFD model (\ref{eq:RFD}). In the same manner, 
replacing the closed diffusive term (\ref{eq:diffusive_term}) of the Lagrangian passive scalar gradient 
equation (\ref{eq:Lag_psi}) yields the RFD model of the passive scalar (\ref{eq:PSRFD}).

\subsection{The detailed derivations of the gradients of the
drifts} \label{Appendix2} To compute the gradient of the
exponential terms of $V\left(\A \right)$ and $M \left(\psi, \A
\right)$ with respect tensor $\A$, it needs to be extended. Up to
the second order of $\tau$, the power series of the matrix
exponential $e^{\textbf{X}} = \sum_{n = 0}^{\infty}
\frac{\textbf{X}^n}{n!}$ is used for the stationary Cauchy-Green
tensor $ \C^{-1} $. Notice that it still possesses the physical
features of the full drifts of this model
\cite{moriconi-pereira-grigorio:2014}. The expansion process is
ordered in the following points: \begin{itemize} \item The power
series of the matrix exponential to extend $\C^{-1} $
gives: \begin{align*} \C^{-1} & = \left( e^{\tau \A }\,\, e^{\tau
\A^{\text{T}}} \right)^{-1} \\ & = \left(\sum_{n = 0}^{\infty}
\frac{\left(-\tau \A^{\text{T}} \right)^n}{n!}\right) \left(\sum_{n =
0}^{\infty} \frac{\left(-\tau \A \right)^n}{n!}\right) \\ & =
\mathbb{I} - \tau \, ( \A + \A^{\text{T}} ) + \frac{\tau^2}{2} \left( \A^2 +
2 \, \A^{\text{T}} \A + \left(\A^{\text{T}} \right)^2 \right) + O \left(\tau^3
\right).  \end{align*} Then, the trace of $\C^{-1}$ after the
truncation to the second order
is \begin{equation} \label{eq:TrC_trunc} \Tr \left(\C^{-1}\right) =
3 + \tau^2 \, \Tr \left(\A^2 \right) + \tau^2 \, \Tr \left(\A^{\text{T}} \A
\right), \end{equation}

where the linearity property of the trace operator and the fact that $\Tr \left(\A \right) = \Tr  \left(\A^{\text{T}} \right) = 0$ 
(due to   incompressibility) and $\Tr \left(\A^2 \right) = \Tr \left( \left(\A^{\text{T}} \right)^2 \right)$ are used. 

	\item Substituting the expanded version of $\C^{-1}$ and its trace
	in the drift term (\ref{eq:detA})
	gives: \begin{equation} \begin{split} V \left( \A \right)& = -
	\A^2 + \frac{\Tr (\A^2 )}{3 + \tau^2 \, \Tr \left(\A^2\right) +
	\tau^2 \, \Tr \left(\A^{\text{T}} \A \right)} \,\, \biggl[ \mathbb{I} - \tau
	\, \left( \A + \A^{\text{T}} \right) + \\ & \,\,\,\,\,\,\, \frac{\tau^2}{2}
	\biggl( \A^2 + 2 \, \A^{\text{T}} \A + \left( \A^{\text{T}} \right)^2 \biggr) \biggr]
	- \frac{\A }{3} \ \biggl[3 + \tau^2 \, \Tr \left(\A^2\right) +
	\tau^2 \, \Tr \left(\A^{\text{T}} \A \right)
	\biggr].  \label{eq:Vtruc} \end{split} \end{equation} The quantity
	$ 1 / (3 + \tau^2 \, \Tr \left(\A^2\right) + \tau^2 \, \Tr
	\left(\A^{\text{T}} \A \right) )$ can be rewritten in terms of Maclaurin
	series as follows, $$ \frac{1}{3 \left(1-x\right)} = \frac{1}{3}
	\sum_{n=0}^\infty x^n, \ \ \ \ \ x \coloneqq - \frac{1}{3} \biggl(
	\tau^2 \, \Tr \left(\A^2 \right) + \tau^2 \, \Tr \left(\A^{\text{T}} \A
	\right) \biggr). $$ Thus, $$\frac{1}{3 + \tau^2 \, \Tr
	\left(\A^2\right) + \tau^2 \, \Tr \left(\A^{\text{T}} \A \right)} =
	\frac{1}{3} - \frac{1}{9} \biggl( \tau^2 \, \Tr \left(\A^2\right)
	+ \tau^2 \, \Tr \left(\A^{\text{T}} \A \right) \biggr) + O\left(\tau^3
	\right).$$ \item Inserting the last equality into equation
	\eqref{eq:Vtruc} and considering only the second order terms of
	$\tau$ yields the truncation formula of $V \left(\A \right) $
	(\ref{eq:detA}),~\cite{grigorio-bouchet-pereira-etal:2017}:
	
\begin{equation}
\label{eq:truncated_V}
 V \left(\A \right) = \sum_{p=1}^{4} V_{p}\left( \A \right),
\end{equation}
where $V_p \left(\A \right)$ contains all the components of $O\left(\A^p \right)$, that is:

$$V_{1}\left(\A \right) = - \A,$$

$$V_{2} \left(\A \right)  = - \A^2 + \frac{\mathbb{I}}{3}\Tr ( \A^2 ),$$

$$V_{3} \left(\A \right) = - \frac{\tau}{3} \, \Tr(\A^2 )  \left( \A+ \A^{\text{T}} \right) - \frac{\tau^2}{3} \, \A \, \left(  \,\Tr \left( \A^2 \right) 
+ \,\Tr \left( \A^{\text{T}} \A \right) \right),$$

\begin{equation*}
\begin{split}
V_{4} \left(\A \right)  & = - \frac{\tau^2}{9} \,  \Tr (\A^2)\,  \mathbb{I} \,  \left( \Tr (\A^2)+ \Tr (\A^{\text{T}} \A ) \right) \\
& \,\,\,\,\,\,\,\, + \frac{\tau^2}{6} \,  \Tr (\A^2) \left(\A^2 + 2 \, \A^{\text{T}} \, \A  + \left(\A^{\text{T}} \right)^2 \right).
\end{split}
\end{equation*}

\item Similarly, the extension version of $M \left(\psi, \A \right)$, resulting from substituting the truncated 
trace (\ref{eq:TrC_trunc}) into the drift of the PS-RFD (\ref{eq:det_theta}), is
\begin{equation}
\label{eq:truncated_b}
M   \left(\psi, \A \right) = - \A^{\text{T}} \, \psi - \frac{1}{3 \, T_\theta} \biggl(  3 + \tau^2 \, \Tr \left(\A^2\right) + \tau^2 \, 
\Tr \left(\A^{\text{T}} \A \right) \biggr) \,\, \psi.
\end{equation} 
\end{itemize}

Now, obtaining the gradient tensors $\nabla_{A_{ij}} V\left(\A\right)_{kl}$, $\ \nabla_{A_{ij}} M\left(\psi, \A \right)_{k}$ 
and $\ \nabla_{\psi_{k}} M\left(\psi, \A\right)_{n}$ (required for instanton equations (\ref{eq:Inst_eqs}) ) from the truncated drifts (\ref{eq:truncated_V}, \ref{eq:truncated_b}) is straightforward computations, as shown:
\begin{itemize} 
	\item The first gradient tensor is 
	\begin{equation}
	\begin{split}
	\left( \nabla_{\A} V \left(\A\right) \right )_{klij} =  \nabla_{A_{ij}} V  \left(\A\right)_{kl} = \sum_{p=1}^{4} \nabla_{A_{ij}} V_{p}\left(\A\right)_{kl},
	\end{split}  \label{Grad_tilda_V}
	\end{equation}
	where,
	$$\nabla_{A_{ij}} V_{1}  \left(\A\right)_{kl}  = - \frac{\partial A_{kl}}{\partial A_{ij}} = - \delta_{ki} \delta_{lj},$$	
	\begin{align*}
	\nabla_{A_{ij}} V_{2}  \left(\A\right)_{kl} & = \frac{\partial}{\partial A_{ij}} \left[ - A_{kl}^2 + \frac{1}{3} \, \delta_{kl} \, \Tr (\A^2 ) \right] \\
	& = - \delta_{ki} \, A_{jl} - A_{ki} \, \delta_{lj} + \frac{2}{3} \, \delta_{kl} \, A_{ji},
	\end{align*}
	\begin{equation*} 
	\begin{split}
	\nabla_{A_{ij}} V_{3}  \left(\A\right)_{kl} & = - \frac{\tau}{3} \frac{\partial}{\partial A_{ij}} \biggl[ \Tr (\A^2 ) 
	\left(A_{kl}+ A_{lk} \right) + \tau \, A_{kl} \left( \Tr (\A^2 )  + \Tr (\A^{\text{T}} \A ) \,  \right) \biggr] \\
	& = - \frac{\tau}{3} \left(  2 \, A_{ji}  \left( A_{kl} + A_{lk} \right)  + \Tr (\A^2 ) \, \left(\delta_{ki} \, \delta_{lj} + \delta_{li} \, \delta_{kj} \right) \right) \\
	& - \frac{\tau^2}{3} \left( \delta_{ki} \, \delta_{lj}  \left( \Tr (\A^2 )  + \Tr (\A^{\text{T}} \A ) \,  \right)  + 2 A_{kl}   \left(  \, A_{ji}  +  \, A_{ij} \right)  \right),
	\end{split}
	\end{equation*}
	\begin{equation*}
	\begin{split}
	\nabla_{A_{ij}} V_{4}  \left(\A\right)_{kl}
	= \frac{\partial}{\partial A_{ij}} \biggl[ 
	& - \frac{\tau^2}{9} \, \Tr (\A^2) \, \delta_{kl}   \, \left( \Tr (\A^2) + \Tr (\A^{\text{T}} \, \A)  \right) \\
	& + \frac{\tau^2}{6} \, \Tr (\A^2) \, \left( A^2_{kl} + 2 \, A_{mk} \, A_{ml} + A^2_{lk} \right) \biggr] \\
	& = - \frac{2}{9} \, \tau^2 \, \delta_{kl} \, \biggl[ A_{ji}  \, \left(  2 \, \Tr (\A^2) +  \Tr (\A^{\text{T}} \A )  \, \right) + \Tr(\A^2) \, A_{ij}  \biggr] \\
	& + \frac{\tau^2}{3} \biggl[ A_{ji} \, \left( A^2_{kl} + 2 A_{mk} \, A_{ml} \,  + A^2_{lk}  \right) + \Tr (\A^2) \, \biggl( \, \delta_{kj}  \, A_{il} \,  + A_{ik} \, \delta_{lj}  \\
	& +\frac{1}{2} \left(  \delta_{ki} \, A_{jl} + A_{ki} \, \delta_{lj} + \delta_{li} \, A_{jk} + A_{li} \, \delta_{kj} \right) \biggr) \biggr].
	\end{split}
	\end{equation*}
	The following relations are used:
	$$ \frac{\partial A^2_{kl}}{\partial A_{ij}} = \delta_{ki} \, A_{jl} + A_{ki} \, \delta_{lj}, \ \ \ \ 
	\frac{\partial \, \Tr (\A^2 )}{\partial A_{ij}} =  A_{ji} + A_{ji} = 2 A_{ji}, \ \ \ \
	\frac{\partial \, \Tr (\A^{\text{T}} \A)}{\partial A_{ij}} = 2 A_{ij}. $$
		
	\item The second gradient tensor is 
	\begin{equation}
	\begin{split} 
	\left( \nabla_{\A} M \left(\psi, \A \right) \right)_{kij} = \nabla_{A_{ij}} M  \left(\psi, \A \right)_{k} & = \frac{\partial}{\partial A_{ij}} \biggl[ - A_{mk} \, \psi_m \\
	& - \frac{1}{3 \, T_{\theta }} \biggl( 3 + \tau^2 \, \Tr(\A^2) + \tau^2 \, \Tr (\A^{\text{T}} \A )  \biggr) \,\, \psi_k \biggr] \\
	& =  - \delta_{kj} \psi_i - \frac{2 \,  \tau^2}{3 \, T_{\theta }}\left(  A_{ji} + A_{ij} \right) \,\, \psi_k.  
	\label{Grad_tilda_b}
	\end{split}
	\end{equation}
	\item The third gradient tensor is 
	\begin{equation}
	\begin{split}
	\left(\nabla_{\psi} M \left(\psi, \A \right) \right)_{nk} = \nabla_{\psi_{k}} M  \left(\psi, \A \right)_{n} & = \frac{\partial}{\partial \psi_{k}} \left[- A_{mn} \, \psi_m
	- \frac{1}{3 \, T_{\theta }} \biggl( 3 + \tau^2 \, \Tr(\A^2) + \tau^2 \, \Tr (\A^{\text{T}} \A )  \biggr) \,\, \psi_k \right] \\
	& = - A_{kn} - \frac{1}{3  \, T_{\theta }} \biggl( 3 + \tau^2 \, \Tr(\A^2) + \tau^2 \, \Tr (\A^{\text{T}} \A )  \biggr) \,\, \delta_{nk}. \label{Grad_tilda_b_wrt_psi}
	\end{split}
	\end{equation}
\end{itemize}

\end{document}